# COMPARATIVE STUDY OF CONGESTION CONTROL TECHNIQUES IN HIGH SPEED NETWORKS

SHAKEEL AHMAD[1, 2], ADLI MUSTAFA[1], BASHIR AHMAD[2], ARJAMAND BANO[3] AND AL-SAMMARRAIE HOSAM[4]

[1]*School of Mathematical Sciences, University Sains Malaysia (USM) Penang Malaysia*
[2]*ICIT, Gomal University, D.I.Khan, Pakistan*
[2]*School of Mathematical Sciences, University Sains Malaysia (USM) Penang Malaysia*
[3]*Mathematics department, Gomal University, D.I.Khan, Pakistan*
[4]*Center for IT and Multimedia, Universiti Sains Malaysia, Penang, Malaysia*
shakeel_1965@yahoo.com, adli@cs.usm.my, bashahmad@gmail.com, hossamjanwer@yahoo.com

**Abstract:** **Congestion in network occurs due to exceed in aggregate demand as compared to the accessible capacity of the resources. Network congestion will increase as network speed increases and new effective congestion control methods are needed, especially to handle "bursty" traffic of today's very high speed networks. Since late 90's numerous schemes i.e. [1]-[10] etc. have been proposed. This paper concentrates on comparative study of the different congestion control schemes based on some key performance metrics. An effort has been made to judge the performance of Maximum Entropy (ME) based solution for a steady state GE/GE/1/N censored queues with partial buffer sharing scheme against these key performance metrics.**

## I. INTRODUCTION

The Internet and wireless technologies are growing rapidly and have been a tremendous success in the past few years. Its presence in every day life is a fact. Traditional slow speed networks have been forced to merge with the high speed networks. But due to increase in Internet size and no. of users, clients are likely to experience longer delay, more packet loss and other performance degradation issues because of network congestion. Formally this problem was tackled by network service providers in terms of keeping utilization of the network low, which may regard as an infeasible solution. As the Internet is gradually dominated by the IP and packet switching, so to increase the network performance in terms of satisfactory level of service to clients is considered as challenging problem [11]. In today's Internet end systems, congestion control mechanism is performed at transports layer.

Because of delay sensitive nature of multimedia applications, they need to be operated on priority basis for satisfying required quality of service constraints [12]. Network traffic produced by these multimedia applications is known to be sensitive in nature and because of random queueing in routers there is a chance of occurrence of delay jitters and end to end delay. In most of the congestion control mechanisms, network routers are equipped with tail drop mechanism having finite capacity queue. When the server is busy, tail drop mechanism accommodate the incoming packets temporarily but upon queue full stage the arriving packets are dropped accordingly.

Apart from simplicity, the technique may suffer various problems i.e. lockout behaviour, global synchronization and full queue [13]. The problem of full queue is the main problem which can produce longer delay and make this mechanism an inappropriate choice for real time applications.

## II. PRELIMINARIES

Before taking into consideration our research work, it would be helpful to re-examine the close connection among type of network traffic, network congestion and buffering in network routers. Internet is defined as the network of networks connected by means of routers. Router directs the packets across the links with the help of bidirectional links. The router decides on the basis of information obtained from routing table about the next ongoing destination link to which packets move to. Line card attached to each of these links are used to perform the packet processing job like stripping off the packet header to make routing decisions. In real time network environment capacity of each link is finite and aggregate demand as compared to the available capacity of the resources may exceed. So the moment, link exceeds its available capacity known to be an overloaded and when this happens it becomes congested. This congestion may be persistent (permanent) or transient (temporary). In case of transient congestion packet arrived abruptly in burst. In transient case solution to congestion is possible by providing a considerable buffer space in router for allowing packets for out-bound link to spend short period before being forwarded to next link. In case of persistent congestion, to avoid from packets drop due to full buffer one possible solution is to increase the size of buffer space with increase in length of the congested period but increase in buffer space however is not an ultimate solution.

Two popular approaches used to control congestion in Internet routers are:

- Congestion prevention, which comes to play before network faces congestion in this case the end systems need to negotiate with the network so that no more traffic than the desired quantity, the network can handle, will be allowed into the network therefore no congestion will occur. This case is also known as *"Open-Loop Congestion Control"* because when the initial negotiation is made between router and the end-system after that both systems will act independently and as a result the end-system get no information from the network about the current traffic and network status, therefore termed as *"Open-Loop Congestion Control"*.

- The $2^{nd}$ approach on the other hand comes into play after the network faces congestion, most of the end-systems in today's networks use reliable data transport protocol such as transmission control protocol (TCP) [3], which has an ability to recognize congestion indicators i.e. lost packets and responding to congestion by reducing the transmission rate. This type of congestion is also termed as "Closed Loop Congestion Control" since the end-system needs to get feedback information from the network about the current congestion status. In this case end-system responds to congestion signal by reducing the load it generates and tries to match the available capacity of the network in order to alleviate the congestion status. We called this type of congestion control method as *"closed-loop"*.

The data transfer between end systems in packet oriented network such as Internet occurs in shape of fixed and variable units of packets of limited size. In general packet oriented networks get congested locally therefore congestion control mechanism usually perform to improve network overall performance and hence it is achieved by controlling the load produced by the network traffic. Based on the current load condition of the network, the congestion control is done through controlling the sending rate of data streams of each source which not only used to prevent congestion but also leads to high utilization of the available band width.

Network protocol frequently inform the sending sources about the current load conditions of the network and as a result the sources store these load conditions in the congestion control variable and these variables are accordingly used for controlling the congestion which leads to achieve high bandwidth utilization and better performance. But this approach has serious limitation i.e. additional overhead is required by the congestion control information that is transferred through the network protocol.

In addition the network protocol and routers are not directly involved to control the congestion in network where as the protocols working on top of the network protocol are responsible to control the congestion and in this case each source on the basis of information stored in its congestion control variable locally perform the activity of congestion control but the major problem with this approach is that the network information collected by the sender does not reveal the fresh status of the network which leads to sub optimal congestion control in terms of overall network performance and utilization.

The remaining paper is organized as follows: Section 2 presents major performance measures and an overview of subset of congestion control schemes. Conclusion is presented in section 3.

## III MAJOR PERFORMANCE MEASURES & OVERVIEW OF CONGESTION CONTROL SCHEMES

The major performance metrics under consideration are:

- Throughput
- Mean Queue length
- Packet Loss Probability
- Link Utilization
- End-to-End Delay or Latency

The most widely deployed congestion control mechanisms are:

### Drop Tail

Drop tail is the simplest and most widely used congestion control scheme in the current Internet routers. It works on first-in-first out (FIFO) based queue of limited size, which simply drops any incoming packets when the queue becomes full. Because of its simple nature, it's easy to implement. Apart from simplicity other advantages include suitability to heterogeneity and its decentralized nature moreover its FIFO based queue provides better link utilization and it helps to absorb the bursty traffic.

Drop tail-TCP Reno routers, have two major drawbacks as pointed out in Braden et al. [22] i.e. its **lock-out** behaviour and the **full queue** phenomena. The lock–out behaviour involves monopolizing of available bandwidth by a single or a few sources which is usually the result of global synchronization [23, 24]. The problem of full-queue is a serious one and it refers to the situation when queue becomes full (or almost full) for long periods of time, consequently which results large end-to-end delays. The other schemes like Drop front on full or Random drop on full drop packets proportional to buffer share's flows hence solving the problem of Drop Tail lock out behaviour but these schemes are unable to solve the problem of full queues. To overcome these problems, one of the possible solutions is to detect congestion earlier and then accordingly to acknowledge the sources about congestion through congestion notification before queue gets overflow. The mechanisms who adopt this strategy are known as "**Active Queue Management (AQM) Schemes".**

We describe such solutions below in terms of Random Early Detection algorithm and its different variants.

For given performance measures Drop Tail behave as:

- **Throughput:** Full buffer state implies low throughput.

- **Mean Queue length:** N.A

- **Packet Loss Probability:** Upon full queue state packets start to drop.

- **Link Utilization:** Provides better link utilization in case of small queue.

- **End-to-End Delay or Latency:** Implies high end-to-end delay in case of full queue.

### AIMD: Additive Increase/Multiplicative-Decrease

In traditional TCP, the feed back control algorithm used to avoid congestion is the "additive increase/multiplicative-decrease (AIMD)". This algorithm is basically used to implement TCP window adjustment as described in [33]. When congestion takes place, AIMD linearly expended congestion window with exponential decrease in it.

The general rule of additive increase is to increase the congestion window by 1 maximum segment size (MSS) every round trip time (RTT) up to the detection of packet loss. Upon detection of packet loss, there is multiplicative decrease in window size i.e. to cut the congestion window to half of the size. The packet loss event is represented either by event of receiving three (3) duplicate acknowledgements or timeout event.

Other related algorithms for fairness in congestion control are MIAD, MIMD and AIAD.

Furthermore in TCP-Reno and TCP-Tahoe, when they are in congestion avoidance phase, the process of additive increase is adopted same as that of AIMD. But in case of packet drop in TCP-Tahoe, more conservative policy is used instead of multiplicative decrease .i.e. protocol enters again into the slow start phase by resetting its congestion window. Whereas in TCP-Reno, upon receiving 3 DACKS by senders multiplicative decrease have been observed in both window and SSThreshold. In spite of the fact that it exhibits fair behaviour with bulk data transfer but it has few limitations [34] i.e. that all flows have the same RTT and the network response arrives at the same time to all users, even when they have the same RTT.

- **Throughput:** Upon congestion detection reduces window size by half.

- **Mean Queue length:** It should not be empty for significant time.

- **Packet Loss Probability:** Packets are dropped when aggregate transmission rate of active connections exceeds the network capacity.

- **Link Utilization:** In case of small queue size link utilization may be decreased due to back off action.

- **End-to-End Delay or Latency:** End-to-end delay increases because of large queue size.

### DECbit Mechanism

DECbit is one of the earliest examples used to control the congestion at routers [25]. The bit in packet header to control congestion in this mechanism is know as congestion indication bit and it is used to provide feedback to the sources for controlling flow of traffic accordingly. In this mechanism, routers set congestion indication bit in arriving packet headers when mean queue length (MQL) exceeds value of 1. This mechanism in general uses windows based flow control for controlling the traffic flow and windows of data packets are updated upon once every round trip times (RTT). The sources decreased window size exponentially in case half of the packets in last window had the congestion indication bit set; otherwise they increased size of the window linearly.

The limitation of this algorithm lies in its averaging queue size mechanism for a limited time period.

For given performance measures DECbit mechanism behaves as:

- **Throughput:** Incremental throughput gained for applying extra load on network is small.

- **Mean Queue length:** It oscillates between empty to non empty state.

- **Packet Loss Probability:** In case of behaving sources packet loss probability can be reduced considerably.

- **Link Utilization:** considerably good.

- **End-to-End Delay or Latency:** It can be reduced by keeping MQL close to 1.

### Random Early Detection (RED):

RED algorithm for RED Gateways was first of all proposed by Sally Floyd and Van Jacobson [5], it calculates the average queue size by using a low pass filter with Exponential Weighted Moving Average (EWMA). RED addresses the shortcomings of traditional Drop Tail algorithm. Router using RED signals incipient congestion to TCP by dropping packets probabilistically before the queue becomes full and this drop probability is depending on running average queue size ($q_a$). If the average queue size is between minimum threshold ($min_{th}$) and maximum threshold ($max_{th}$) the packet is marked or dropped with some probability, $p_a$ which is given as:

$$p_a \leftarrow p_b / (1 - count \cdot p_b)$$

*Where* $p_b \leftarrow \max_p (q_a - \min_{th}) / (\max_{th} - \min_{th})$

If $q_a > max_{th}$, then the packet is dropped. If $q_a < max_{th}$, then the packet is forwarded through.

When $q_a$ oscillates between minimum threshold ($min_{th}$) and maximum threshold ($max_{th}$), $p_b$ varies linearly between 0 and $max_p$.

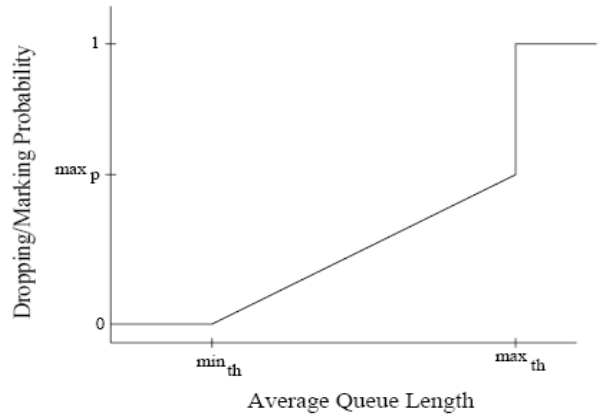

Figure-1: The marking/dropping behavior of RED [36]

Although RED is probably the most widely used AQM scheme for congestion avoidance and control but it has been observed from various studies [14], [15], [16], [17] that the performance of RED is highly dependent upon the environment where it is used as well as the way its parameters are tuned. Thus, the performance benefits of RED as claimed in [5] and others are not mostly true.

- **Throughput:** Depends upon traffic intensity and the mode its parameters are adjusted.
- **Mean Queue length:** Packet drop probability increases with increase in mean queue length.
- **Packet Loss Probability:** When $q_a > max_{th}$, then the packets start to drop.
- **Link Utilization:** Link utilization is efficient in case of small queue/buffer size.
- **End-to-End Delay or Latency:** Delay may increase in case of large queue size.

**RED Variants:**

**Adaptive RED (ARED):**

ARED is a variant of RED [18] used to change the parameters of RED adaptively according to the observed traffic load. ARED algorithm works by inferring whether or not RED should become more or less aggressive by observing the behavior of mean queue length ($q_a$). ARED algorithm accordingly adjusts the value of $max_p$ by examining the behavior of mean queue length. Opposing to the RED algorithm, ARED marking function changes depending on the setting of $max_p$ i.e. in case of low congestion the marking/dropping probabilities remain also low until mean queue length reaches $max_{th}$ whereas in case of high congestion the marking/dropping probabilities grow quickly as the mean queue length exceeds $min_{th}$. Main advantage of ARED lies in its automatic setting of parameters in response to the change in network traffic load and its drawback lies in a fact that ARED is not clear to decide about best and optimum policy of parameter change.

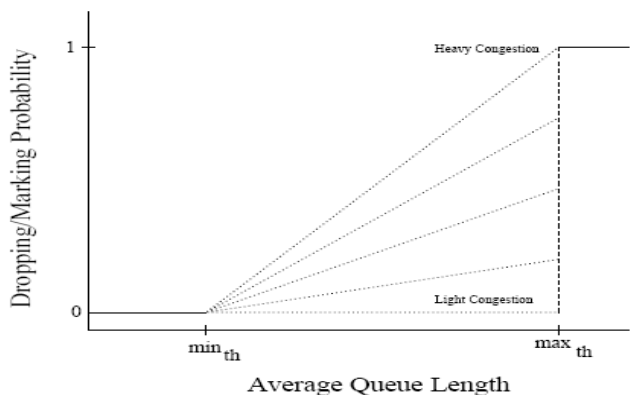

Figure-2: The marking/dropping behavior of ARED [36]

- **Throughput:** Maintain high throughput levels across all workloads.
- **Mean Queue length:** $max_p$ depends on mean queue length value.
- **Packet Loss Probability:** Maintain low packet loss rates across all workloads.
- **Link Utilization:** In order to prevent under utilization less aggressive early detection is required when small numbers of flows are active.
- **End-to-End Delay or Latency:** Same as the case with RED .i.e. delay may increase in case of large buffer size.

**Explicit Congestion Notification (ECN):**
The Explicit Congestion Notification extends RED in a way that instead of dropping a packet it just marks it when the average queue size ($q_a$) lies between minimum threshold ($min_{th}$) and maximum threshold ($max_{th}$) [19]. ECN requires both end-to-end and network support [20]. The receiver upon receipt of packet with congestion bit set acknowledged the sender about incipient congestion which onward activates the congestion avoidance algorithm at the source end accordingly. ECN can't be relied upon completely towards elimination of packet losses as indications of congestion. Also it does not eliminate the need for fast retransmit & retransmit timeout mechanisms for detecting dropped packets. Moreover ECN requires changes to TCP and IP header.
For given performance measures ECN behaves as:

- **Throughput:** TCP/ECN, the average aggregated throughput is quite higher than without ECN.
- **Mean Queue length:** MQL examines the congestion level and when queue becomes full packets are last in burst.
- **Packet Loss Probability:** In case of TCP connection, Packet loss probability is low. As TCP's are sensitive to even a single packet loss.
- **Link Utilization:** Better link utilization can be achieved by TCP/ECN.

- **End-to-End Delay or Latency:** Reduced end-to-end delay in TCP/ECN.

**Blue:**
Blue is another extension of RED developed by Wu-Chang and Feng et al. [21] which uses packet loss and link utilization (rather than queue size) as a control variables to measure the network congestion.
The algorithm used in BLUE works as:
If the queue size exceeds a threshold $L_{th}$, the packet marking probability $P_m$ is increased by a set rate $r_1$, and in case of idle link it is decreased by $r_2$.
When the queue becomes greater then $L_{th}$, $P_m$ increases by $r_1$ for every freeze_time ($F_t$) seconds.
When the link is idle, $P_m$ decreases by $r_2$ for every freeze_time ($F_t$) seconds.
Freeze_etime is minimum time interval between two successive updates of $P_m$.
Hence the control variable $P_m$ used to control the arrival rate and maintains the buffer below threshold thus most of the time link not remains idle. Overall the algorithm tries to minimize packet loss rate and helps to keep the buffer stable. The major drawback of BLUE is the fact that it is not scalable.

- **Throughput:** Maintains high throughput levels.
- **Mean Queue length:** Maintains small queue length.
- **Packet Loss Probability:** Provide low packet loss rate.
- **Link Utilization:** Link utilization is considerably high.
- **End-to-End Delay or Latency:** Maintains less buffer size results in low end-to-end delay.

Brief overview of some other variants of RED is described as below:

**Stabilized RED (SRED):**
Ott et al. in [26] developed another variant of RED in terms of SRED. The algorithm matches an arriving packet with that chooses on random basis. On successful match a "hit" occurs and sequences of hits not only used to estimate no. of active connection but also used to find potential candidate for non behaving sources. In this mechanism blocking or packet loss probability depends on no. of active connections and current size of a queue.

**Flow RED (FRED)**

FRED algorithm was developed by Lin and Morris [27]. They "On the basis of simulation results they concluded that RED is unfair towards multiple class type traffic and this is because of the fact that regardless of bandwidth used by all flows RED at any given time imposes same loss rate on these flows". FRED algorithm continuously monitors the current queue occupancy by a given flow. If a large size of queue is consistently occupied by some flow then after detection algorithm bounds it to small queue size in order to maintain the fairness among different flows.

**Dynamic RED (DRED):**

J. Aweya et al. proposed another variant of RED in [28] and its main objective is to maintain the queue length near to threshold value defined by the user. DRED make use of control theory to adopt the packet dropping probability. DRED plus points lie in its independence from the no. of flows that passes through the router and bounded delay.

**Gentle RED (GRED):**

In RED if qa>maxth all the arriving packet are dropped accordingly which leads to RED oscillatory behaviour. Floyd introduces gentle RED in [29] to reduce the undesired oscillations in buffer size. To increase the throughput and to smooth the probability of dropping he made use of the maximum dropping probability (.1) between the two thresholds and the maximum dropping probability is 1 when the MQL reaches double the maximum threshold as shown by the following figure:

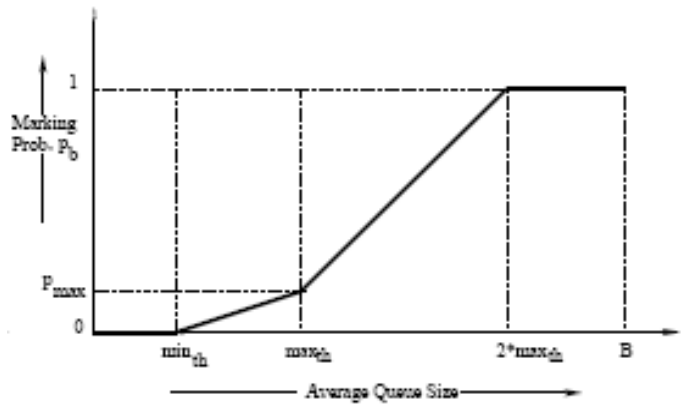

Figure-3: The marking/dropping behavior of GRED [37]

**Random Exponential Marking (REM):**

Snajeewa Athuraliya et al. proposed another variant of RED in terms of REM in [32] with aims to achieve high link utilization, negligible packet loss & end-to-end delay and scalability. The key features of REM as described in [32] are:

- Match rate clear buffer attempts to match user rates to capacity of the network while clearing or stabilizing buffers around a small target regardless of the no. of users.
- Sum of link prices depends on end-to-end marking/dropping probability.

The main drawback of REM is that it gives no incentives to behaving sources.

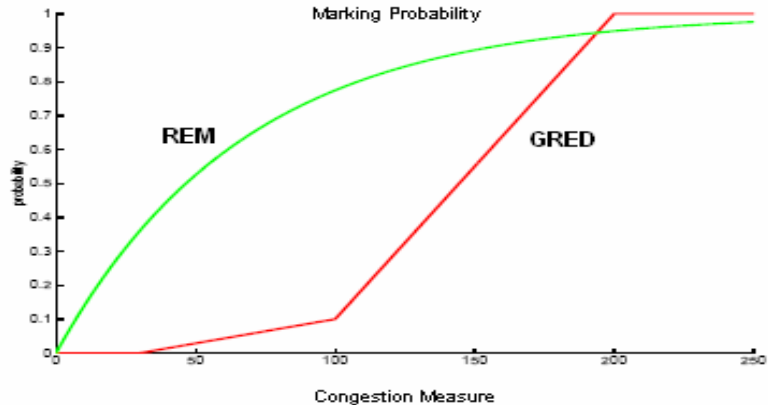

Figure-4: Marking probability of REM and GRED [38]

The other prominent RED variants are DSRED, FDRED, SFB, CBT-RED etc. whereas CRED and PRED are explicitly designed for ATM switches.

Summary of the various RED variants can be represented by the following table:

|  | Drop Function | Control Variable | Changes from original RED |
|---|---|---|---|
| FRED | Single linear | Per-flow queue length | Per-flow queue length, number of active flow |
| FBRED | Single linear | Average queue length | Per-flow $Max_{drop}$ |
| SRED | 3 segment step | instantaneous queue length and number of active flow | Step drop function, number of active flows, instantaneous queue |
| CBT-RED | Single linear | Average queue length | Class based threshold |
| XRED | Single linear | Average queue length | Priority based drop |
| BRED | 4 segment step | Per-flow queue length and number of active flows | Per-flow queue length, number of active flows, step drop function |
| DSRED | Two linear | Average queue length | Two linear drop function with different slope, |
| BLUE | Step function | Link utilization and packet loss | Step increase/decrease function, link rate, packet loss |
| REM | Exponential function | Link rate mismatch and buffer difference | Exponential function, link rate mismatch and buffer difference |
| SFB | Step function | Instantaneous queue length | Organize sub-queue in Bloom filter |

Table-1:Summary of RED variants [30]

## Partial Buffer Sharing (PBS) Based Congestion Control Mechanism [35]:

Partial buffer sharing scheme plays an important role towards effective congestion control mechanism in network routers. This scheme effectively controls the allocation of buffer to various traffic classes according to their delay constraints. The motivation behind this scheme is to meet the diverse demands of QoS which can be achieved by improving the loss performance of the high priority traffic while degrading the performance of the low priority traffic.

**PBS Algorithm** works as follows:

*Step−1 : Set descending sequence of threshold to $N_i$ ($N_i > 0$, $i = 1, 2, …, R$)*

   *corresponding to Q with finite capacity & single server having R priority classes*

*Step−2 : To meet desired QoS demands adjust threshold values under different load conditions accordingly*

*Step−3 : The highest priority jobs of class 1 can join the queue subject to space availability in Q*

*Step−4 : Jobs with lower priority class i ($i = 2, …, R$) can only join the Q if total jobs in Q, $N < N_i$ ($N_i ≤ N_{i-1}$)*

*Step−5 : Once the number of jobs waiting for service reaches $N_i$*

   *All jobs with lower priority will be lost on arrival But higher priority jobs can still join the queue*

   *Until it reaches threshold value, $N_j$ ($j = 1, …, i−1$)*

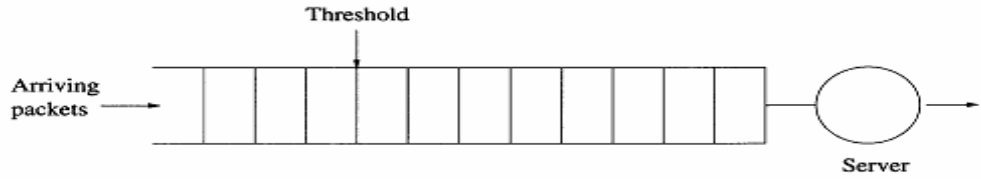

Figure 5: Arrangement of a simple threshold based PBS scheme [35]

The above arrangement shown by Figure 5 assumes a buffer with single FIFO queue.

Solution based on Maximum Entropy Methodology for a stable GE/GE/1/N censored queue with a single server, finite capacity and multiple classes of traffic under PBS scheme in terms of the joint aggregate ME queue length distribution $\{P(n), n \in \Omega \}$ is given by [35]:

$$P(0) = \frac{1}{Z}$$

$$P(k) = \sum_{i=1}^{R} \Pr ob\left(Q_{i;k}\right)$$

$$= \frac{1}{Z}\left(\prod_{j=1}^{R} x_j^{k_j}\right) \sum_{j=1}^{R} k_j \left(\frac{\left(\sum_{i=1}^{R} k_i - N_j\right)!}{\prod_{i=1}^{R}\left(k_i - N_j\right)!}\right) g_j y_j^{\delta(k)}$$

The external bursty traffic and service time have been modeled using the generalised exponential (GE) distribution.

The blocking or packet loss probability is given as [35]:

$$\pi_i = \sum_{k=0}^{N} \delta_i\left(k\right)\left(1 - \sigma_i\right)^{\left[N_i - k\right]^+} P_{N_i}\left(k\right)$$

Where

$$\delta_i\left(k\right) = \begin{cases} \dfrac{r_i}{r_i\left(1 - \sigma_i\right) + \sigma_i}, & k = 0 \\ \\ 1, & \text{otherwise} \end{cases}$$

For given performance measures it behaves as:

- **Throughput:** Figure-6 shows that throughput values for delay sensitive traffic and delay tolerant streams vary by increasing the threshold position. Both curves coincide before the entire buffer becomes shared which shows bursty nature traffic.

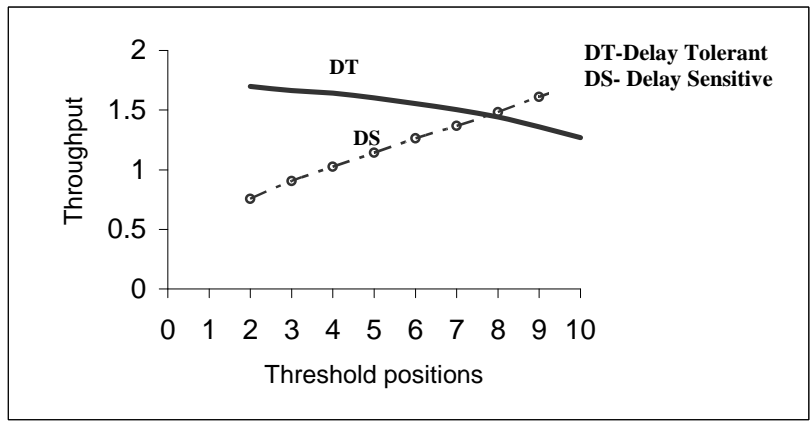

**Figure-6:** Effect of threshold positions on throughput

- **Mean Queue length:** Mean queue length size plays an important role towards allocation of threshold positions.

- **Packet Loss Probability:** Figure-7 shows that by increasing the position of threshold the packet loss probability for delay sensitive traffic decreases whereas it increases in case of delay tolerant traffic.

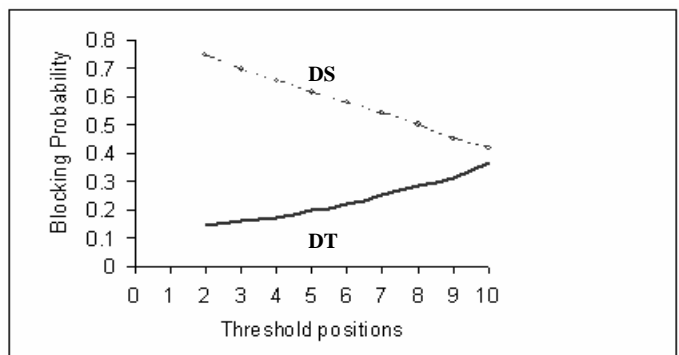

**Figure-7:** Effect of threshold positions on blocking probabilities

- **Link Utilization:** Link utilization increases with increase in threshold positions.

- **Mean Response Time:** Figure-8 shows that mean response time increases with increase in threshold positions both in case of DS and DT traffic streams.

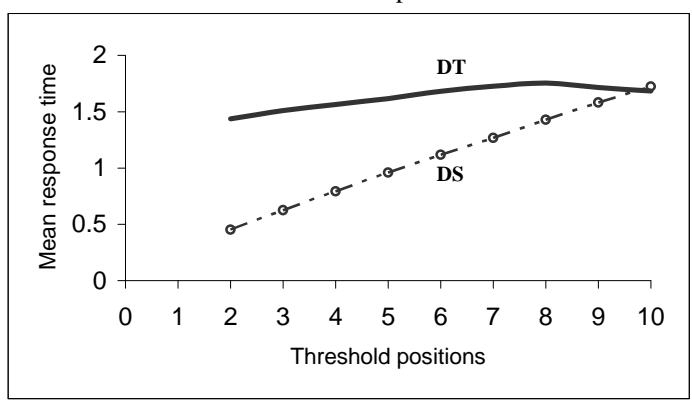

**Figure 8:** Effect of threshold positions on mean response time

## IV. CONCLUSION

This paper briefly surveys comparative analysis of different congestion control algorithms on the basis of some key performance measures. It is observed and concluded that at present no single congestion control mechanism can solve all of the problems due to the wide number of parameters that have impact on system's performance. In addition it is also concluded that in today's high speed network, the nature of congestion is not really known and one can't easily characterise the different levels of congestion along with the facts that what is an extreme condition of congestion, how long does it lost and what is the percentage of dropped packets???. Thus more research is needed in this area of networking.